\documentclass[twocolumn,showpacs,showkeys,superscriptaddress,preprintnumbers,amsmath,amssymb]{revtex4}
\usepackage{graphicx}
\usepackage{dcolumn}
\usepackage{bm}
\usepackage{amsmath}
\usepackage{enumitem}
\usepackage{color}
\begin{document}

\title{Collisional Model for Granular Impact Dynamics}

\author{Abram H. Clark}
\affiliation{Department of Physics and Center for Nonlinear and Complex Systems, Duke University, Durham, North Carolina 27708, USA}
\author{Alec J. Petersen}
\affiliation{Department of Physics and Center for Nonlinear and Complex Systems, Duke University, Durham, North Carolina 27708, USA}
\author{Robert P. Behringer}
\affiliation{Department of Physics and Center for Nonlinear and Complex Systems, Duke University, Durham, North Carolina 27708, USA}

\begin{abstract}
When an intruder strikes a granular material from above, the grains
exert a stopping force which decelerates and stops the intruder. Many
previous studies have used a macroscopic force law, including a drag
force which is quadratic in velocity, to characterize the decelerating
force on the intruder. However, the microscopic origins of the force
law terms are still a subject of debate. Here, drawing from previous
experiments with photoelastic particles, we present a model which
describes the velocity-squared force in terms of repeated collisions
with clusters of grains. From our high speed photoelastic data, we
infer that `clusters' correspond to segments of the strong force
network that are excited by the advancing intruder. The
model predicts a scaling relation for the velocity-squared drag force
that accounts for the intruder shape. Additionally, we show that the
collisional model predicts an instability to rotations, which depends on the intruder
shape. To test this model, we perform a comprehensive
experimental study of the dynamics of two-dimensional granular impacts
on beds of photoelastic disks, with different profiles for the leading
edge of the intruder. We particularly focus on a simple and useful
case for testing shape effects by using triangular-nosed intruders. We
show that the collisional model effectively captures the dynamics of
intruder deceleration and rotation; i.e., these two dynamical effects
can be described as two different manifestations of the same
grain-scale physical processes.
\end{abstract}
\date{\today}

\keywords{Granular materials, Granular flow, Impact}
\pacs{47.57.Gc, 81.05.Rm, 78.20.hb} 

\maketitle
\section{Introduction} Granular impact has been studied extensively
for many years (e.g.,
\cite{Euler1745,Poncelet1829,Allen1957,Forrestal1992,Tsimring2005,Katsuragi2007,
  Goldman2008,Goldman2010,Takehara2010,Clark2012,Clark2013,Clark2013-2,Ciamarra2004,Ambroso2005,Bruyn2004,Walsh2003,Nelson2008,Newhall2003},
and references therein), as understanding how the intruder's momentum
is collectively transferred to the grains and its energy dissipated
has obvious practical applications in ballistics, meteor impacts, and
industrial processes. In general, the dynamics depend on the
microscale physical characteristics of the system (e.g., type of
grains, initial packing fraction, presence of an interstitial fluid,
etc.) and the intruder velocity (e.g., faster or slower than the
granular sound speed). A complete description represents a complex
problem in granular rheology.

In the present study, we consider impacts with initial intruder speeds
that are well below the granular sound speed, but fast enough
that static or quasi-static effects provide only a modest portion of
the decelerating forces. The intruders impinge on a moderately
compacted, dry granular bed. For this case, many previous studies
\cite{Euler1745,Poncelet1829,Allen1957,Forrestal1992,Tsimring2005,Katsuragi2007,
  Goldman2008,Goldman2010,Takehara2010,Clark2012,Clark2013,Clark2013-2,Seguin2009} have
successfully used a macroscopic force law to describe the `slow'
dynamics, i.e., over time scales that are much slower than granular
fluctuation time scales \cite{Clark2012,Clark2013-2}. These force laws typically
contain a velocity-squared drag force which dominates the bulk of the
deceleration. In addition, they contain other terms that typically are
important late in the collision process, when static granular forces
that support the weight of the intruder become important.

Here, we base our discussion on a typical model (see also previous
work \cite{Clark2012,Clark2013,Clark2013-2}) with the form:
\begin{equation}
F=m\ddot{z}=mg-f(z)-h(z)\dot{z}^2.
\label{eqn:forcelaw}
\end{equation}
$F$ is the force on the intruder, $z$ is the depth within the
material, $z = 0$ corresponds to the point of initial contact between
the intruder and the unperturbed granular surface, $mg$ is the
gravitational force, $f(z)$ is a static term, $h(z)$ characterizes the
strength of the inertial ($v^2$) term, and dots denote time
derivatives. Often, $h(z)$ is assumed to be constant, though we find
that it can have an initial transient. Although such models are
empirical, they typically capture the `slow' dynamics of intruder
trajectories. However, the grain scale origins of these terms are
still poorly understood, and the focus of this study is to improve
this understanding.

The velocity-squared term, $h(z)\dot{z}^2$, is typically understood as
an inertial term, which models dynamic momentum (and energy) transfer
from the intruder to the particles. In a previous study, using
photoelastic particles \cite{Clark2012}, we showed that, although the
slow or average dynamics are well captured by
Eq.~\eqref{eqn:forcelaw}, the force is not smooth on faster time
scales. Rather, the force is spatio-temporally highly fluctuating due
to intermittent emission of acoustic energy along relatively
long-lived granular networks which are excited locally along the
intruder-granular interface. In other experiments using circular- and
elliptical-nosed intruders \cite{Clark2013}, we also showed that
changes in intruder shape had a strong effect on $h(z)$. It is this
effect that we pursue here, first by considering other shapes for the
intruder, and second by constructing a collisional model that
explicitly involves the intruder shape.

\begin{figure*}[th!] 
\centering
 \includegraphics[clip,trim=35mm 0mm 35mm 3mm,width=\textwidth]{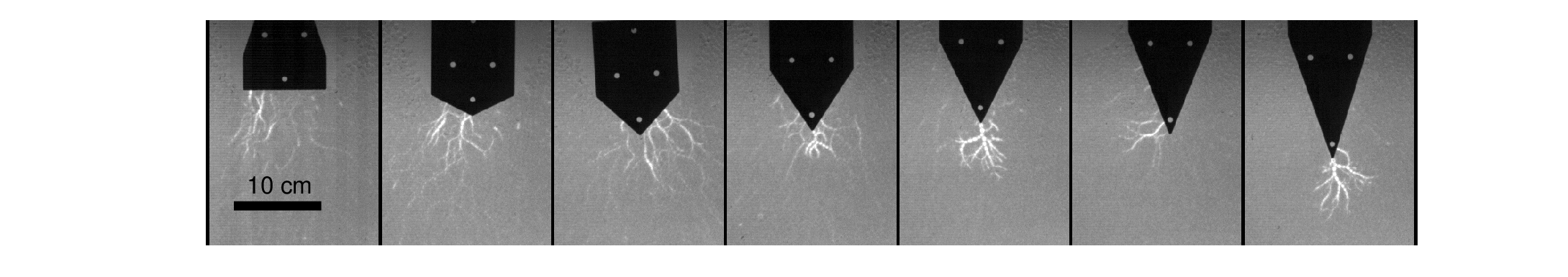}
 \caption{Still frames showing each of the seven triangular-nosed
   intruders (described in the text) with leading edge slopes of $s=0$
   to $s=3$, from left to right. These images are chosen during times
   when the intruders have collided with and excited networks. It is
   the networks that form the clusters of particles discussed in the
   model. Here, grains that are carrying an instantaneously large force
   appear bright. Note that the force networks or force chains are
   oriented roughly normally to the intruder surface at the
   intruder-granular interface. Additionally, the collisions shown for
   the $s=$1.5, 2, and 3 intruders occur at the intruder tip, which
   illustrates the large forces that occur there when $s$ is large.}
 \label{fig:frames}
\end{figure*}

Specifically, in this paper, we present a comprehensive study of the
dynamics of intruders with triangular noses of varying shape but
constant mass and width, including the deceleration and rotation of
the intruder. These data augment earlier results
\cite{Clark2012,Clark2013} for circular and ogive intruders. We fit
the dynamics of deceleration to the force law in
Eq.~\eqref{eqn:forcelaw}, and thereby measure $f(z)$ and $h(z)$ for
varying intruder shape. As a way to understand the shape dependence of
$h(z)$ and the dynamics of the rotations, we propose a mesoscopic
collisional model, where the intruder is decelerated through random,
repeated collisions with `clusters' of grains. By clusters, we refer
to force-chain-like networks that are acoustically excited as in
Fig.~\ref{fig:frames}. This collisional model contains a
velocity-squared drag force, $h(z)\dot{z}^2$, which depends on the
intruder size and shape in a way that incorporates the interactions of
the intruder with these structures. The drag force includes an unknown
$O(1)$ multiplicative constant that is common to the $h(z)$
coefficient for all intruders. Once this coefficient is determined
using one shape of intruder, all other aspects of the model can easily
be tested by using other intruder shapes. Using this approach, we
find good agreement between the experimental data and the model for
the velocity-dependent drag force. In these experiments, we consider
a range of intruder types, including intruders with triangular
`noses', as well as intruders with curved shapes such as ogives and
disks that we have studied previously~\cite{Clark2013}. Additionally,
we show that the collisional model captures the rotational dynamics of
the intruder, which become more striking as the length of the
intruder nose increases.

The key feature for the inertial term of the model is momentum
transfer to clusters of grains, based on nominally specular inelastic
collisions of clusters with the intruder. These collisions are
affected by the local shape and speed of the intruder at the collision
point. We also find that the tip of each triangular intruder (which is
slightly rounded to avoid breaking particles) yields a
disproportionately large collisional effect, compared to other parts
of the intruder. By adding the tip contribution to the rest of the
shape-dependent contribution, we show that the collisional model
captures the velocity-squared drag for all intruders.

The static term, $f(z)$, in Eq.~\ref{eqn:forcelaw} is also of
interest, although the collisional term dominates the stopping
process. However, in the present experiments, we fix the cross
sectional area and mass of the intruder, which are likely to affect
$f(z)$. We then observe that even large variation in shape, at fixed
width and mass, does not substantially affect $f(z)$. However, a
recent study \cite{Durian2013} suggested this term was due to static,
depth-dependent friction, and had a strong dependence on intruder
shape, especially while the intruder nose was only partially
submerged. Understanding how $f(z)$ depends on intruder shape should
also give insight into the grain-scale processes which control it.
Brzinski, et al. \cite{Durian2013} placed intruders of various shapes
(spheres, cylinders, and cones) at various depths in a granular bed.
They then measured for each depth the maximum force which the granular
material could support before failing. Additionally, they imposed a
controlled airflow through the grains to alter the the strength of
hydrostatic pressure via gravitational loading. They focused on the
regime \textit{before} the intruder nose was fully submerged. In this
regime, they found shape dependence which is consistent with a static,
depth-dependent pressure which points normally inward at each
grain-intruder contact, where the static pressure is a direct result
of hydrostatic loading from the grains above. Once the intruder nose
was fully submerged, they found the same static force for all
intruders with the same cross-sectional area ($f(z)= kz$, where $k\sim
\pi R^2$), with a small correction for the nose shape. They then
inferred that this force is indeed the static term in dynamic
experiments, $f(z)$, by measuring the final penetration depth versus
drop height for spheres and cylinders. They found that these results
were consistent with their previous measurements of the static
force. In our experiments, which focus on the regime after the nose is
submerged, we observe that even large variation in shape (at fixed
intruder width) does not substantially affect the static force term,
$f(z)$, a result which is consistent with the findings of Brzinski, et
al. as well as other slow-drag studies \cite{Albert2001,Goldsmith2013}. 
Also, our measurements (here, and in \cite{Clark2013}), as well
as measurements by Goldman and Umbanhowar \cite{Goldman2008}, show a
substantial offset term once the nose is fully submerged,
$f(z)=f_0+kz$, where $f_0$ scales with the intruder mass. 

The outline of the remainder of this paper is as follows. In Section
II, we present the details of the collisional model, including
predictions for the drag force and torque, which we will reference
throughout the remainder of the paper. In Section III, we describe the
experiment and the techniques used to collect and analyze the data. In
Section IV, we present the data for the intruder trajectories
(including depth, velocity, deceleration, and rotations, as well as
$f(z)$ and $h(z)$ for each intruder), and we then compare the
experimental results to the predictions of the collisional model for
the velocity-squared drag force and rotational dynamics.
Figs.~\ref{fig:hvsI} and \ref{fig:GammaInstability} are the
culmination of this analysis. Section V contains a summary,
conclusions, and outlook.

\section{Collisional Model} 
\subsection{Assumptions}

The basis of the model proposed here is that the intruder transfers
momentum to the granular material through a sequence of random
collisions that excite the force network, as in Fig.~\ref{fig:frames}.
To develop this idea further, we consider quantitative grain-scale
force response to impacts, reported earlier \cite{Clark2012} . These
results were obtained using particles made from a `hard' photoelastic
material, such that the speed of the intruder was always slow compared
to the sound speed through the material. For these conditions, the
intruder deceleration was dominated by interactions with filamentary
networks of grains (often referred to as force chains) that carried
relatively large forces. As the intruder pushed through the granular
bed, acoustic pulses generated at the intruder-granular interface,
propagated along these strong force networks, carrying momentum and
energy away from the intruder. These networks changed with time, but
persisted long enough to carry one or more acoustic pulses. In
particular, they were long enough lived to exhibit clear tracks on
space-time plots of the local granular force~\cite{Clark2012}. We refer
to the grains in one of the networks as a cluster. But, we emphasize
the filamentary quasi-1D structure formed by such a
cluster. Specifically, clusters are not usually collections of grains
in a simple (e.g. roughly circular) 2D region.

\begin{figure}
\includegraphics[width=0.9\columnwidth]{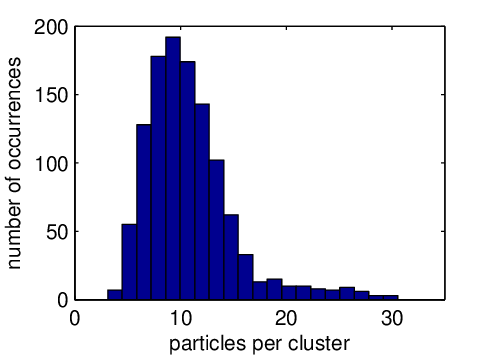}
\caption{(color online.) Histogram showing the average cluster size, which we estimate by dividing the total number of bright pixels within $1.5$ intruder radii of the bottom edge of a circular intruder (although the measurement is insensitive to the size of the region used) by the number of bright pixels within one particle diameter of the intruder nose. The above plot was obtained from a single trajectory using a circular intruder, and the plot only includes frames where at least 50 pixels are bright at the intruder edge, to make sure that the intruder is in contact with a particle cluster. This result is typical for all trajectories.}
\label{fig:clustersize}
\end{figure}

Typically, the networks emanated from the lower boundary of the
intruder at angles that were close to the boundary normal. Thus,
collisions transfered momentum into the intruder at a set of
point-like contacts, such that the momentum carried along the network
is normal to the local surface (at least initially). Typical images in
Figure~\ref{fig:frames} demonstrate this feature and correspond to one
frame for each of the seven triangular intruders. Each frame is chosen
at a time for which there is a collision with a grain cluster. Since
we cannot resolve particles, we are unable to definitively say how
many particles are involved in each one of these clusters. However, as
a way to estimate the cluster size, we divide the total number of
bright pixels beneath the intruder ($1.5$ intruder radii from the
bottom edge of a circular intruder) by the total number of bright pixels within one particle
diameter from the bottom edge of the intruder, using only image frames
where at least one particle (about 50 pixels) is bright at the
intruder edge. A pixel is denoted `bright' if it is a fixed amount brighter
than the background intensity. The threshold is chosen as 10\% greater than the background intensity,
which conforms well with what one identifies as `bright' by eye. Figure~\ref{fig:clustersize} shows a histogram of this
measurement from a single trajectory using a circular intruder, which
shows an average cluster size of about ten particles.

\begin{figure}[th!]
\includegraphics[width=0.6\columnwidth]{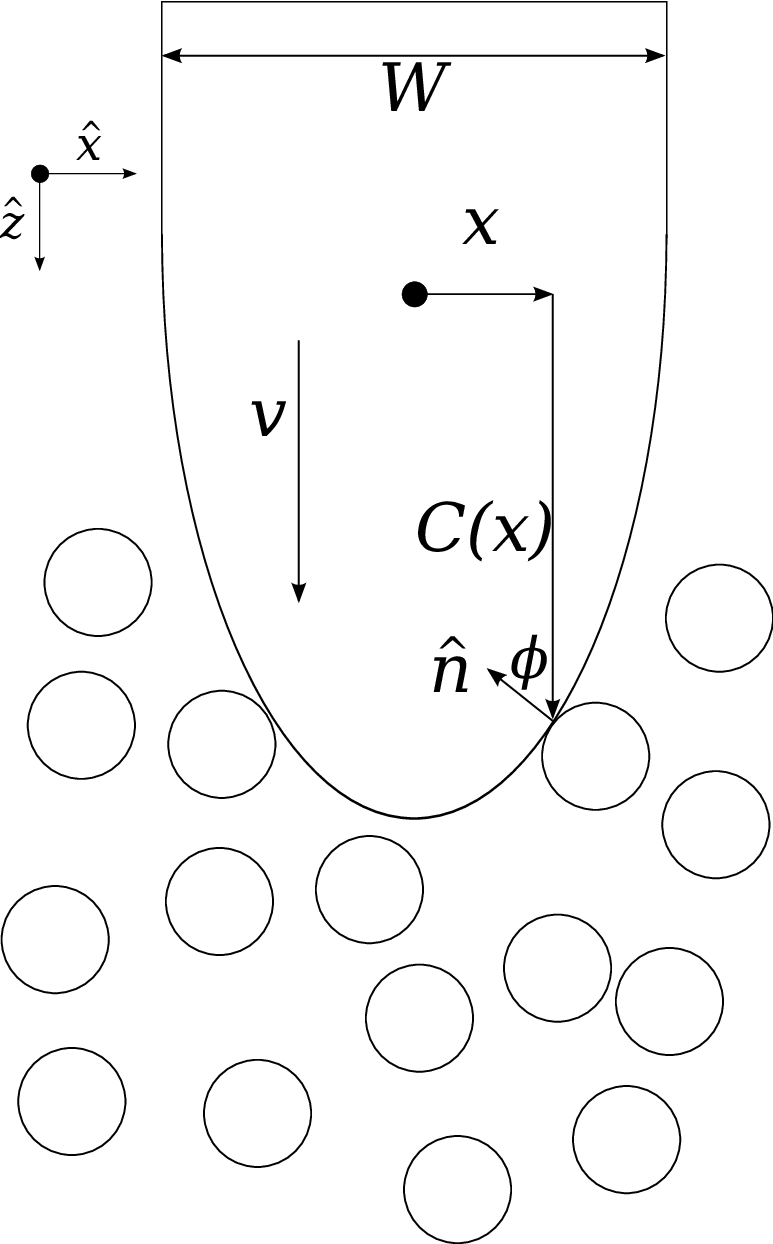}
\caption{Sketch of the collisional model, where an intruder of width
  $W$ randomly collides with grain clusters (represented by the open
  circles) as it moves at velocity $v$. The collisions occur along the
  `nose', i.e. the leading edge of the intruder, at positions
  $\vec{r}=x\hat{x}+C(x)\hat{z}$ measured from the center of mass.
  These collisions involve momentum transfer normally into the
  intruder, along normal vector $\hat{n}$.}
\label{fig:cartoon}
\end{figure}

We then construct a simplified version of this process in order to
calculate the force acting on the intruder. As in
Fig.~\ref{fig:cartoon}, we imagine an object with mass $m$, profile
specified by $C(x)$, and width $W$, moving at a velocity $v$,
undergoing repeated inelastic collisions with small, stationary
objects. Here, these small objects are grain ``clusters", in the above
sense, which we take to have mass $m_c$. We expect that $m_c$ is
greater than the mass of a single grain, since a collision involves
force-chain-like structures that remain in contact over a finite
time. In Fig.~\ref{fig:cartoon}, the clusters are now represented as
mesoscopic `particles'. A related collisional picture was proposed by
Takehara, et al. \cite{Takehara2010}. In their study, an intruder was
subjected to constant-velocity drag through a granular medium, and the
drag force was measured with a high-speed force sensor. The drag force
was measured to be quadratic in velocity, and, based on
momentum-transfer considerations as well as images of the motion of
the grains, Takehara and colleagues argued that it is the result of
repeated collisions with particle clusters which are larger than the
mass of a single grain. Although they could not visualize grain scale
forces, they nevertheless concluded that ``the formation of the
dynamical force chains plays a crucial role."  Our approach is
similar, but we are able to directly verify the role of the granular
network, and relate the structure of that network to the collisional
properties of our intruders. It is also interesting that the speeds
used by Takehara et al. were roughly an order of magnitude smaller
than our fastest impact speeds. This means that the collisional forces
are important over a surprisingly large range of intruder speeds.

We next assume that the clusters behave as quasi-particles which
collide inelastically with the surface of the intruder. We assume
collisions in the direction of the surface normal, $\hat{n}$, that are
inelastic and captured (in that direction) by a restitution
coefficient, $e$. Momentum in the direction parallel to the intruder
interface is unaffected by the collision. A collision imparts momentum
$\Delta \vec{p} =\hat{n}(1+e)\frac{m_c m}{m_c+m}v\cos\phi$, where
$\phi$ is the angle between the velocity and $\hat{n}$. We take the
typical collision time to be $\Delta t =\gamma d/(v\cos\phi)$, where
$d$ is the particle diameter and $\gamma$ is an $O(1)$ constant. Thus,
the average force at a particular location along the intruder surface
is in the $\hat{n}$-direction, given by:
\begin{equation}
\vec{f}=\frac{\Delta\vec{p}}{\Delta t}=\frac{(1+e) v^2\cos^2\phi}{\gamma d}\left(\frac{m_cm}{m_c+m}\right)\hat{n}.
\label{eqn:collforce}
\end{equation}
Note that, since the bronze intruders we use are much more dense than the photoelastic material used for particles, the reduced mass in our case is approximately equal to the mass of a cluster, $\frac{m_cm}{m_c+m}\approx m_c$. We assume that collisions are equally likely per unit normal
area. Thus, the relative number of collisions in a length of intruder
surface $dl$ is $dn=\beta dl/d$ (where $\beta$ is another $O(1)$
parameter), so $d\vec{F}=\hat{n}f\,dn=(\hat{n}f\beta/d)dl$. If the
shape of the intruder is given by $z=C(x)$, as in
Fig.~\ref{fig:cartoon}, then $dl=(1+C'^2)^{1/2}dx$, and
$\cos\phi=(1+C'^2)^{-1/2}$.

This force is quadratic in the velocity and depends on the local shape
of the intruder surface, varying as $\cos^2 \phi$. By integrating this
force over the intruder surface, we can obtain a specific prediction
about the effect of intruder shape on the $v^2$ term, effectively
giving $h(z)$, in terms of a shape factor, multiplied by an $O(1)$
multiplicative term that is the same for all intruders, regardless of
shape, for a given bed material.

\subsection{Upward Force}

We first consider only vertically upward ($z$-component) of $d\vec{F}$
by integrating over the leading edge of the intruder and keeping only
the $z$-component:
\begin{eqnarray}
F_z&=&\int d\vec{F}\cdot\hat{z}\nonumber \\
&=&\frac{(1+e)\beta m_cm}{\gamma d^2(m_c+m)}\left[\int_{-W/2}^{W/2}dx(1+C'^2)^{-1}\right]v^2\nonumber \\
&=& B_0\cdot I[C(x)]\cdot v^2.
	\label{eqn:zforcecoll}
\end{eqnarray}
The constant $B_0=\frac{(1+e)\beta m_cm}{\gamma d^2(m_c+m)}$ contains
various system parameters which are nominally the same for all intruders. The size and shape effects are contained in
$I[C(x)]$, defined as:
\begin{equation}
I[C(x)]\equiv\int_{-W/2}^{W/2}dx(1+C'^2)^{-1}.
\label{eqn:I}
\end{equation}
For example, $I = W$, the width of the intruder, if $C' = 0$,
corresponding to an intruder with a flat interface. This model
considers only the collisional part of the drag force, and gives no
prediction about the shape effects for the $f(z)$ term. However, this
term tends to be important only as the intruder is coming to rest. In
Section IV, in our discussion of rotations, we return to this issue,
showing that $>80\%$ of the trajectory (in terms of distance) is
dominated by the velocity-squared drag force. Note too, that, although
the real grain clusters near the surface of the intruder are also
moving, their velocity relative to the intruder should scale with the
intruder velocity, so this analysis should still hold up to a constant
scale factor of order unity.

For this study, we particularly focus on intruders with triangular
noses, intended as two-dimensional `cones' as in \cite{Newhall2003},
which have a constant slope, $|C'(x)|=s$, everywhere except at the
tip. This yields a very simple form for $I[C(x)]$, which allows a
straightforward method to separate shape from other effects:
\begin{equation}
I(s)=\frac{W}{1+s^2}.
\label{eqn:shape-eff-triang}
\end{equation}
The constant slope for conical intruders allows us to unambigously
focus on the directional effects assumed in the collisional model.
Additionally, moderate variation in $s$---here, we use seven
triangular-nosed intruders, with slope $s$ varying between 0 and 3, in
increments of 0.5---gives an order of magnitude change in $I(s)$
(i.e., $I(s=0)=W$ and $I(s=3)=W/10$). This provides a sensitive test
of the collisional model.

We also include data for the collisional drag coefficient measured for
circular- and elliptical-nosed intruders, which was presented in a
recent paper \cite{Clark2013}. For circular or elliptical noses,
$I[C(x)]$ has the following form:
\begin{equation}
I(a,b)=\frac{2b}{\left(\frac{a}{b}\right)^2-1}\left(\frac{\left(\frac{a}{b}\right)^2 \arctan\sqrt{\left(\frac{a}{b}\right)^2-1}}{\sqrt{\left(\frac{a}{b}\right)^2-1}}-1\right),
\label{eqn:Iellipse}
\end{equation}
where $a$ and $b$ are the semi-major and semi-minor axes, respectively
(The limit for a circular nose, $a=b=R$, is well defined, specifically
$I(R)=4R/3$.)

\subsection{Rotation and Torques}

Experiments show that intruders can rotate as they move through the
granular medium. If a symmetric intruder is tilted at an angle
$\theta$ from vertical, the forces on either side are generally
different from what they would be if the intruder were vertically
oriented. Most importantly, the forces are not the same on either side
of the intruder, which is particularly evident for triangular-noses,
where the intruder-cluster collision angles clearly differ from side
to side. Such a tilt has several consequences. First, the vertical
force may differ from the corresponding value when the intruder is
vertical; second, there can be a horizontal component of the force;
and third, there may be a non-zero torque on the intruder. It is
possible to correct for the effect of rotations on the vertical force
by recalculating $I$ based on a $C(x)$ that includes the instantaneous
$\theta$ for the intruder (i.e. calculate $C(x,\theta)$), and use this
in the computation of $I$. Note that the vertical force, which is
determined by $I$, must have an extremum at $\theta = 0$, since the
force must be invariant to whether $\theta$ is positive or
negative. The horizontal force and the torque are not subject to the
same symmetry principle, so a small tilt will lead to values of these
quantities that are $O(\theta)$. In the present experiments, the
corrections to the vertical force and the relative magnitude of the
horizontal force are small compared to the unperturbed part of the
vertical force.

However, the same is not generally true for the torque. Particularly
in the late stages of the dynamics, we observe substantial rotations
for some intruders. Thus, the torque on an intruder at an angle
$\theta$ must be calculated in a method similar to
Eq.~\eqref{eqn:zforcecoll}. This yields quantitative predictions for
the dynamics of rotation, including the prediction of rotational
instability. The total torque, $\vec{\tau}$, about the center of mass
of the intruder is given by integrating $\vec{r} \times \vec{f}$ over
all collisions over the intruder surface, similarly to
Eq.~\eqref{eqn:zforcecoll}:
\begin{eqnarray}
\vec{\tau} &=& \int \vec{r}\times\vec{f}\,\,\frac{\beta}{d}dl \nonumber \\
	&=& B_0v^2\int\vec{r}\times \hat{n}\cos^2\!\phi\, dl
\label{eqn:torque-tot}
\end{eqnarray}
Here, $\vec{r}=x\hat{x}+C(x,\theta)\hat{z}$, $\hat{n}=-\sin\phi\hat{x}-\cos\phi\hat{z}$, and $\sin\phi=-C'(1+C'^2)^{-1/2}$, with $dl=(1+C'^2)^{1/2}dx$ and $\cos\phi=(1+C'^2)^{-1/2}$, as before (see the sketch in Fig.~\ref{fig:cartoon}). This yields:
\begin{eqnarray}
\vec{\tau} &=& B_0v^2 \int dx \left(\frac{CC'}{1+C'^2}+\frac{x}{1+C'^2}\right)\hat{y}\nonumber \\
	&=& B_0 J[C(x,\theta)]v^2 \hat{y}.
\label{eqn:torque-tot2}
\end{eqnarray}

\subsection{Tip Effects}

A last issue concerns the concentration of stress at the tip of the
triangular intruders, particularly the more pointed ones. By
examining the photoelastic response near the intruder, we can separate
the relative contributions from collisions with the tip and collisions
elsewhere on the smooth intruder surface. The tip is quite small, with
a radius of curvature that is a fraction of a particle radius. It is
still possible to calculate a shape factor for the tip,
$I[C_{tip}(x)]$, where $C_{tip}(x)$ is the shape of the rounded
tip. However, this analysis is misleading, because the probability of
a collision at the tip is far greater than elsewhere on the smooth
surface of the intruder. By analyzing the photoelastic response near
the tip, and in other regions, as shown in the top of
Fig.~\ref{fig:photodrag}, we find that the tip contribution to the
velocity-squared force is approximately the same for intruders with
prominent tips (i.e., large $s$), but it is considerably larger per
unit area, by about an order of magnitude, than contributions from
collisions away from the tip. The term due to collisions with the tip
must be included separately in the force and torque calculations for
accurate prediction of the velocity-squared drag and off-axis
rotations. 

Since the collisional model is based on spatially averaged
random collisions, we expect it to break down as the spatial scale of the
intruder (i.e., the size of the tip) approaches the system's microscopic length
scale (i.e., the size of a particle). A flat (or gently curved) section of the intruder
undergoes collisions with excited force networks with probability which 
grows proportionately to its area. However, a small tip is able to contact
individual particles and excite a force network with some finite number of particles,
despite its small size. Thus, we expect a sharp tip to be more efficient
per unit area than a gently curved section.

\section{Experimental Techniques}
\subsection{Experimental Apparatus} 
The experimental apparatus is the same as that used in previous
studies \cite{Clark2012,Clark2013}. It consists of a two-dimensional
granular bed of approximately 25,000 bidisperse, hard, photoelastic
disks (diameters of 6~mm and 4.3~mm, thickness of 3~mm, with
approximately equal numbers of each size of particle) confined between
two thick Plexiglas sheets (0.91~m $\times$ 1.22~m $\times$ 1.25~cm)
separated by a thin gap (3.3~mm). The photoelastic disks are cut from
PS-1 material (Vishay Precision Group; bulk density of 1.28~g/cm$^3$,
elastic modulus of 2.5 GPa, and Poisson's ratio of 0.38). This
material has a bulk sound speed of approximately 2000~m/s, and the
granular sound speed for this system is found to be about 300~m/s
\cite{Clark2012}. Intruders, which are machined from bronze sheet
(bulk density of 8.91~g/cm$^3$ and thickness of 2.3~mm), are dropped
from above the layer with an initial orientation that closely
corresponds to $\theta = 0$, and have initial impact speeds up to $v_0
= 6$~m/s. A Photron FASTCAM SA5 records the process at a resolution of
256$\times$584 pixels ($\sim 10$ pixels per $d$), and at 40,000 frames
per second.
\subsection{Intruder Shape}
The triangular-nosed intruders are comprised of a downward-pointing
isosceles triangle, symmetric about the vertical axis, with opening
angle $2\alpha$, attached to a rectangular tail of the same width as
the base of the triangle, $W=9.65$~cm. The noses of these intruders
are clearly evident in Fig.~\ref{fig:frames}. The length of the tail
is varied to keep the total area, $A=0.0107$~m$^2$, and hence, mass,
$m=0.219$~kg, constant for different opening angles of the nose (for
reference, the $s=3$ intruder has a tail which is 3.81~cm long). Thus,
the intruder nose has a constant magnitude slope $s=\tan^{-1}\alpha$,
except at the tip, which is rounded with a radius of about
1.5~mm. Note that this is smaller than the particle radii, which are
2.1~mm and 3~mm. The $s=3$ intruder is turned upside-down and used as
the $s=0$ intruder.

\subsection{Data Processing}
At each frame, we use distinguishing features of the intruder to
locate its center of mass relative to the initial point of impact and
its angular position relative to the vertical direction with errors of
less than 1 pixel (0.5~mm). This yields the intruder trajectory, and
the rotation angle, $\theta$. By discrete differentiation, combined
with a low-pass filter, we obtain the velocity and acceleration. Using
the data for $z(t)$, $v(t)$, and $a(t)$ for many different
trajectories with varying initial velocities, we fit to a force-law
model such as Eq.~\eqref{eqn:forcelaw}. This allows us to
experimentally measure $f(z)$ and $h(z)$ for each intruder. This
process is described in detail in \cite{Clark2013}.

\begin{figure}[th!]
\includegraphics[clip,trim=0mm 18mm 10mm 10mm,width=0.75\columnwidth]{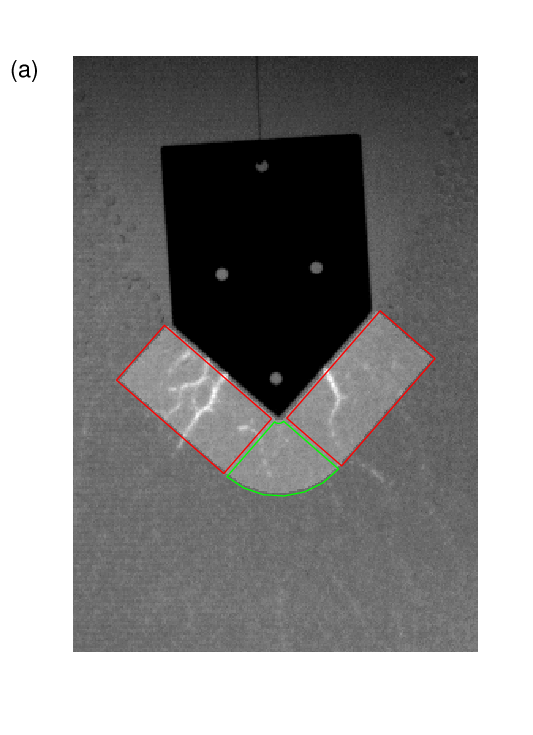}
\includegraphics[clip,trim=0mm 0mm 0mm 0mm,width=0.9\columnwidth]{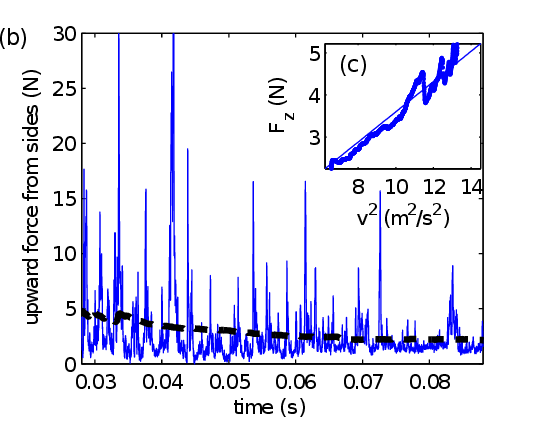}
\caption{(color online.) (a) A photoelastic image showing the
  different regions used to measure the force contributions from the
  sides (outlined in red) and from the tip (outlined in green) of the
  triangular-nosed intruders. (b) Plot of the calibrated
  photoelastic force on the sides of the $s=1$ intruder, for a single
  trajectory with $v_0\approx 3.8$, both at each frame (thin, blue
  line) and after a low-pass filter (thick, black, dashed line). Inset (c)
  shows the low-pass filtered force signal determined photoelastically
  versus $v^2$, where the imposed fit passes through the origin with a
  slope of 0.36, which is the effective drag coefficient contribution
  from the sides of the triangular intruder in this case.}
\label{fig:photodrag}
\end{figure}

Additionally, the photoelastic response on the particles just beneath
the intruder provides another way to measure the force experienced by
the intruder. After removing background-light inhomogenieties, we take
the discrete gradient-squared ($G^2$) of the photoelastic image. As
shown in \cite{Clark2012}, we are able to successfully calibrate $G^2$
as a measure of the force along a particular section of the
intruder. Using the photoelastic response has a distinct advantage: it
allows us to separate the force on different sections of the intruder,
specifically the sides and beneath the tip, as shown in the top panel
of Fig.~\ref{fig:photodrag}. By calibrating $G^2$ per pixel to the
measured local pressure, we can evaluate the force on a given section
of the intruder nose, which we assume points normally inward. This is
supported by the direction of the force chains, which point roughly
normally out of the edge of the intruder nose. The various components
of this force determine the stopping force, the torque, etc. We also
assume that the collisional drag from the tip of triangular-nosed
intruders points straight upward. Finally, by applying a low-pass
filter to the strongly fluctuating time series for $F_z$, and
correlating with $v^2$ for large velocities ($v>2.5$~m/s) where
collisional drag dominates, we measure the collisional drag for that
section of the intruder nose. (We will show that this measurement
agrees well with the measurement of the drag coefficient from tracking
the intruder, which is measured using all velocities, so we believe
that using only fast velocities, with $v>2.5$~m/s, has no effect on
determination of the drag coefficient using the photoelastic
response.) This process is shown in Fig.~\ref{fig:photodrag}, and it
is repeated for approximately ten trajectories per intruder. The
results are then averaged to determine the mean collisional drag for
each region of each intruder.

\section{Comparison of Model to Experimental Data}
\begin{figure}[th!]
\includegraphics[trim=0mm 50mm 0mm 5mm, width=0.9\columnwidth]{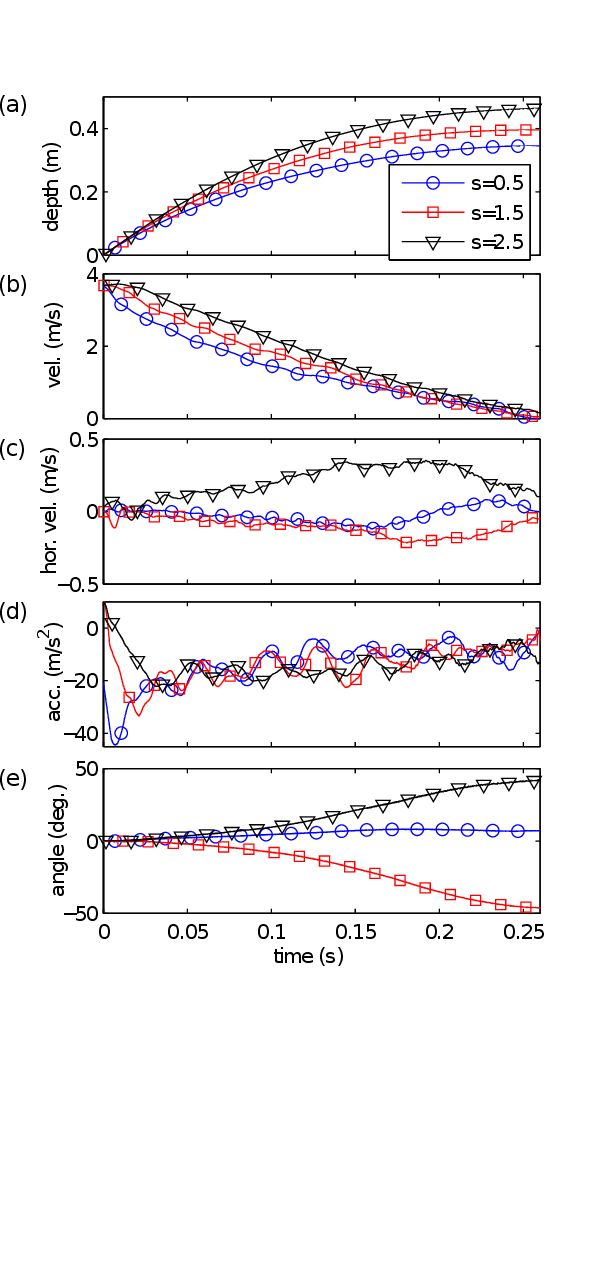}
\caption{(color online.) Plots vs. time of three individual intruder
  trajectories and related information ($s=0.5,1.5,2.5$) with
  $v_0\approx 3.55$~m/s, including (a) depth, (b) downward velocity,
  (c) horizontal velocity, (d) acceleration, and (e) angular orientation of the
  intruder, with initial impact at $t=0$. Many such trajectories are
  used to fit to the force law in Eq.~\eqref{eqn:forcelaw}.}
\label{fig:trajectories}
\end{figure}

\begin{figure}[th!]
\includegraphics[trim=0mm 2mm 0mm 0mm, width=\columnwidth]{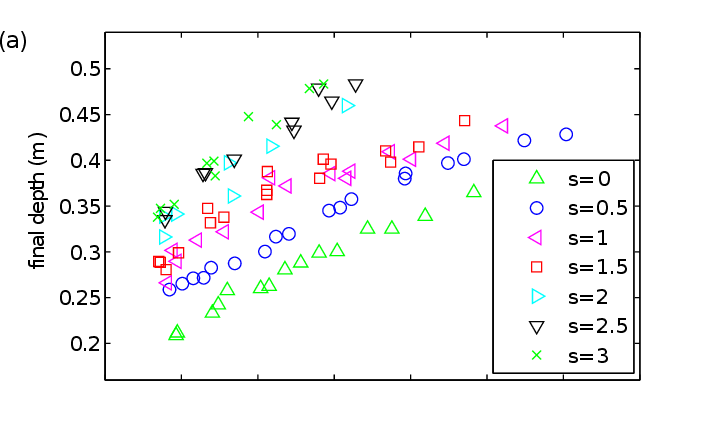}
\includegraphics[trim=0mm 0mm 0mm 6mm, width=\columnwidth]{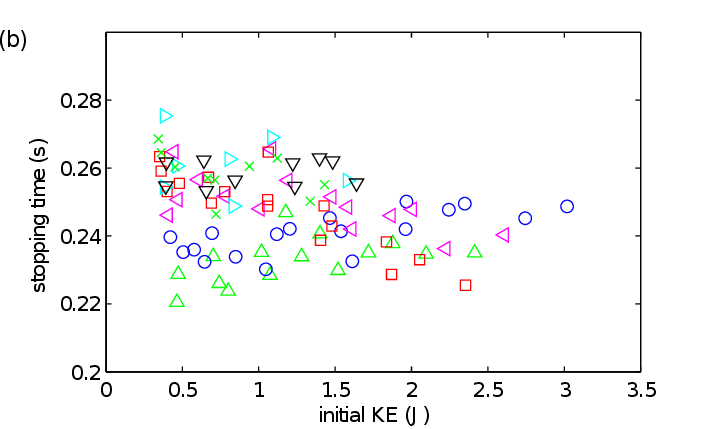}
\caption{(color online.) Plots of final depths (a) and stopping times (b) for all intruders as a function if the kinetic energy at impact. Note that increasing $s$ leads to deeper penetration, but virtually no change in stopping time.}
\label{fig:stop-times-and-depths}
\end{figure}

Figs.~\ref{fig:trajectories} and~\ref{fig:stop-times-and-depths} show
typical impact trajectories for intruders with triangular noses.
Figure \ref{fig:trajectories} shows three trajectories for intruders
with $s=0.5,1.5,2.5$, impacting with similar initial velocities,
$v_0\approx 3.55$~m/s. We show several different quantities: the depth
below the point of initial contact, the downward velocity, the
horizontal velocity, the acceleration, and the angular orientation of
the intruder, with $t=0$ corresponding to initial impact. Increasing
$s$ leads to deeper penetration for the same initial impact velocity,
but the stopping time remains about the same
(Figure~\ref{fig:stop-times-and-depths}). Additionally, increasing $s$
corresponds to a weakening of the initial deceleration at impact,
which is manifest in the form of $h(z)$ for different shapes, as
discussed later. The fluctuations in the acceleration correspond to
fluctuations observed in the photoelastic response \cite{Clark2012};
the particular range of fluctuation frequencies is set by the cutoff
frequency of the low-pass filter used to reduce noise that is
introduced by numerical differentiation of data for the intruder's
location.

Note that the angular rotations for large $s$ can be substantial,
such as in the trajectory from the $s=1.5$ intruder shown in
Fig.~\ref{fig:trajectories}, and that there is no preferred direction
of rotation. This suggests an instability to small perturbations for
the angular orientation of the intruder. The typical intruder
horizontal velocity is quite small compared to the vertical velocity,
so the assumption that the velocity is purely vertical is reasonable.

We now seek to understand shape-dependence of all intruder
trajectories, particularly in terms of the collisional model, as
discussed above. We do this in two parts. First, we consider the
depth, velocity, and acceleration, and fit these data to the force-law
model from Eq.~\eqref{eqn:forcelaw}. By combining this analysis with
the photoelastic analysis, we show that, modulo an overall
normalization factor, the collisional model gives an accurate
prediction of the velocity-squared drag force felt by the intruder.
In particular, by choosing a reasonable value for the normalization,
we find that the shape effect on $h(z)$ is accounted for. Second, we
examine the rotational dynamics, and show that the collisional model
also gives an accurate prediction of these dynamics as well.

\subsection{Force Law Analysis}

\begin{figure}[th!]
\includegraphics[clip,trim=0mm 6mm 0mm 0mm,width=0.9\columnwidth]{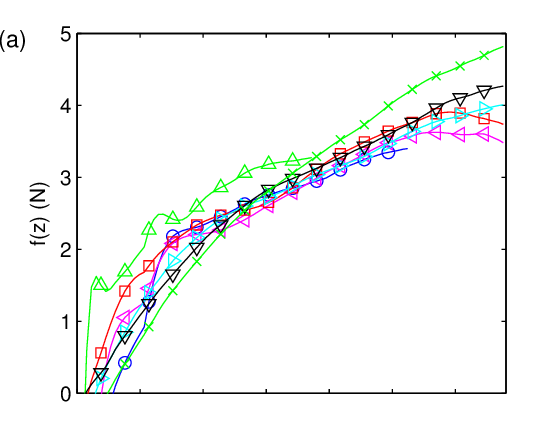}
\includegraphics[clip,trim=0mm 0mm 0mm 4mm,width=0.9\columnwidth]{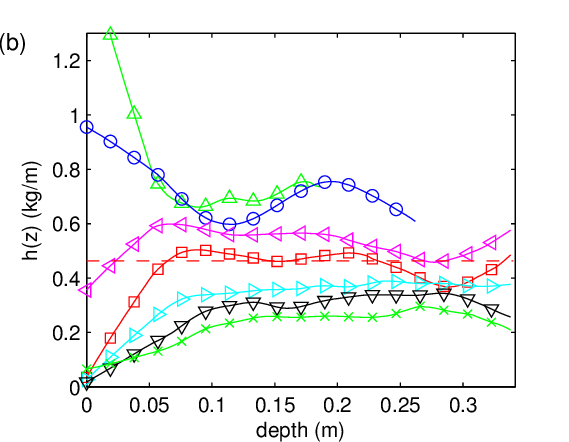} 	      
 \caption{(color online.) 
   Plots of $f(z)$ (a) and $h(z)$ (b) versus depth for 
   each of the seven triangular-nosed intruders, measured from the
   intruder dynamics ($s=0$ to $s=3$ are denoted in order by \textcolor{green}{$\triangle$},
   \textcolor{blue}{$\circ$}, \textcolor{magenta}{$\triangleleft$}, \textcolor{red}{$\square$},
   \textcolor{cyan}{$\triangleright$}, \textcolor{black}{$\triangledown$}, \textcolor{green}{$\times$}, as in Figure \ref{fig:stop-times-and-depths}). Note that $f(z)$ shows almost no
   dependence on intruder shape, and $h(z)$ is relatively constant, denoted $h_0$, after an initial transient
   which is larger or smaller depending on the shape of the intruder
   nose. The red dashed line shows $h_0$ for $s=1.5$.}
 \label{fig:hzcones}
\end{figure}

We first measure $f(z)$ and $h(z)$ for each intruder by using the
depth, velocity, and acceleration data for all trajectories
\cite{Clark2013}. We then ask how these functions, shown in
Fig.~\ref{fig:hzcones}, are affected by the intruder shape, where we
focus particularly on the triangular intruders. The static force term,
$f(z)$, shown in the top panel of Fig.~\ref{fig:hzcones}, is
essentially insensitive to intruder shape, within the scatter of the
data. In this figure, different colors distinguish the various
intruders. By contrast, the collisional term, $h(z)$, shows
significant dependence on shape (bottom panel). This term exhibits an
initial transient, followed by roughly steady-state behavior. We define
$h_0$ to be this steady-state value of $h(z)$ after the initial 
transient. More elongated intruder noses (i.e., larger
$s$) are associated with decreasing $h_0$. The transition from the
initial transient to $h_0$ corresponds roughly to the time at which
the nose penetrates the granular material, and presumably to the
formation of a steady-state velocity field around the intruder. The
transient behavior also depends on the shape, where blunt-nosed
intruders ($s<1$) have $h(z=0)>h_0$, and elongated-nosed intruders
($s\geq 1$) have $h(z=0)<h_0$. Additionally, as discussed previously,
we also measure $h_0$ using the calibrated photoelastic response
starting after the nose is fully submerged. This allows us to examine
the contributions made by different pieces of the intruder nose (i.e.,
the sides or the tip of the triangular nose).

\begin{figure}[h!t]
\includegraphics[clip,trim=0mm 0mm 0mm 0mm,width=0.9\columnwidth]{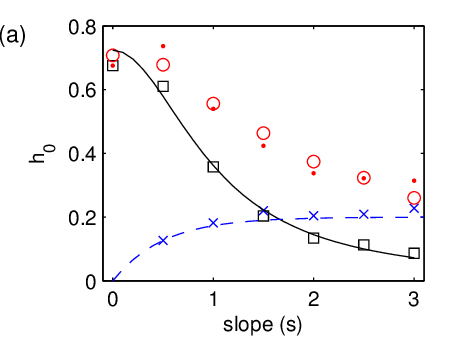}
\includegraphics[width=0.9\columnwidth]{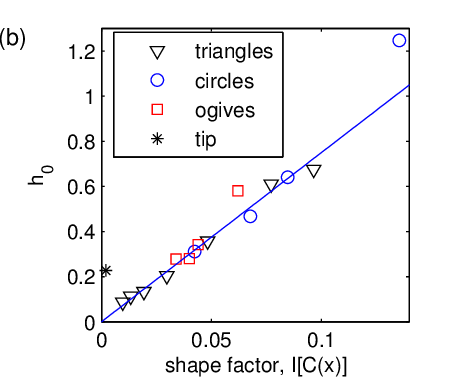}
\caption{(color online.) (a) Plot of $h_0$ measured from the
  intruder acceleration and force measurements inferred from the
  photoelastic response. The values from the intruder acceleration
  (open, red circles), as shown in the main figure, show good
  agreement with the values from the photoelastic signal (red
  dots). The contribution from the tip (blue crosses), measured from
  the photoelastic signal, stays relatively constant for $s>1$. The
  fit line (dashed, blue line) is given by $0.2(1-e^{-2s})$, which is
  an approximate fit assigned by eye. Open,
  black squares show the contribution from the sides, which matches
  extremely well to the model (solid black line),
  $I(s)=W(1+s^2)^{-1}$, as in
  Eq.~\eqref{eqn:shape-eff-triang}. (b) Plot of $h_0$ versus
  $I[C(x)]$ for all intruders. The solid line shows a linear fit
  through the origin, where the slope is
  $\frac{h_0}{I[C(x)]}=B_0\approx 7.6$. The $h_0$ value for triangular
  noses is the photoelastic measurement from the sides (excluding the
  tip), while the asymptotic value (approximately 0.2 N) for the tip
  measurement is shown separately (black asterisk). The collapse
  appears to be robust over a wide range of intruder sizes and shapes,
  with the possible exception of the largest intruder (a circular
  intruder with 20.32~cm diameter), suggesting that the unknown scale
  factors (e.g. $\alpha$, $\beta$) vary as the intruder size or mass
  is changed greatly; see \cite{Clark2013} for further discussion.}
\label{fig:hvsI}
\end{figure}

Figure~\ref{fig:hvsI} shows a summary of the behavior of the
velocity-squared drag coefficient, $h_0$, and its dependence on shape
for different intruders, as well as a comparison to the collisional
model predictions in Eqs.~\eqref{eqn:I} and
\eqref{eqn:shape-eff-triang}. Specifically, the top of
Figure~\ref{fig:hvsI} shows results for triangular-nosed intruders
that demonstrate the two methods of calculating the total collisional
drag coefficient: from the intruder trajectories (open, red circles)
and using the photoelastic response (red dots). These two very
different approaches are in good agreement with each other. Using the
photoelastic data, shown in Figure~\ref{fig:photodrag}, we decompose
these drag coefficients into contributions from the sides (open, black
squares), which matches well to the predictions for the collisional
stresses with the flat surfaces of the triangular-nose intruders
dependence (solid, black line), and contributions from the tip (blue
crosses). These latter data asymptotically approach a constant value
as the aspect ratio is increased from $s=0$ to $s=3$ (i.e., as the tip
is made more prominent), which we capture with an approximate fit of
$h_{tip}\approx 0.2(1-e^{-2s})$. Considering the tip as a small flat
section with a width of 3~mm, its contribution for large $s$ is
approximately 10 times bigger than what occurs for a comparable area
on the flat part of the intruders. The large contribution from the tip
is visually clear by inspection of the photoelastic videos, which show
a surprisingly large amount of acoustic pulses emanating from the
intruder tip. (The examples shown in Figure~\ref{fig:frames} are
representative of this). In this regard, we note that the model is
based on an assumption of locally smooth surfaces such that the flux
of clusters impinging on the surface is a coarse-grained measure. At
the tip, this assumption is violated, implying that the tip needs to
be treated on a separate footing.

In the bottom panel of Figure~\ref{fig:hvsI}, we plot $h_0$ versus
$I[C(x)]$, also including data for circular- and elliptical-nosed
intruders, from \cite{Clark2013}. Since intruders with circular or
elliptical noses are smooth (i.e., have no small tips), we use the
value of $h_0$ extracted from the trajectories. In this figure, we
have also plotted separately the contributions from the sides and from
the tip of triangular-nosed intruders. There is a good linear collapse
of all the data, except perhaps the heaviest circular-nosed intruder
(see \cite{Clark2013} for further discussion). The successful linear
collapse of $h_0$, which is measured from experimental data, versus
$I[C(x)]$, which is calculated from the theory, within an overall
shape-independent constant, is a strong confirmation of the
collisional model and is the first main result of this paper. This
implies that $B_0=h_0/I[C(x)]$, which contains the microscopic details
for collisions, is relatively constant for all intruders used.

\subsection{Rotational Dynamics}
\begin{figure}[t!]
\includegraphics[clip, trim=0mm 0mm 0mm -10mm, width=0.8\columnwidth]{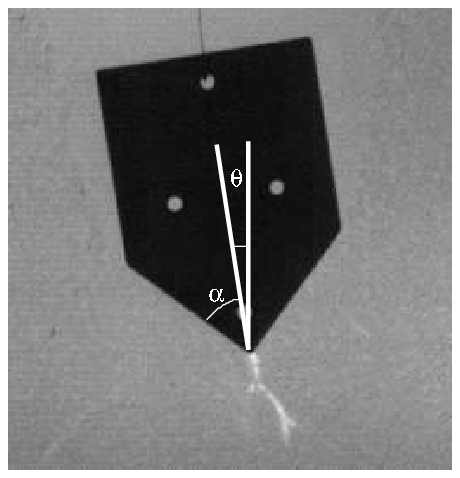}
\caption{Image of a triangular-nosed intruder ($s=1$) which has
  rotated by an angle $\theta$.}
\label{fig:conerotimage}
\end{figure}

Thus far, the collisional model has provided a good description of the
vertical force on the intruder. We now examine the intruder rotations
in the context of the collisional model, as in
Eqs.~\eqref{eqn:torque-tot} and \eqref{eqn:torque-tot2}. An image of
an intruder which is rotated by an angle $\theta$ is shown in
Fig.~\ref{fig:conerotimage}.

For the symmetric intruders used here,when $\theta=0$, the
integral $J[C(x,\theta)]$ is equal to zero (since $x$ and $C'$ are
both odd functions in $x$, and everything else is even). However, as
noted earlier, for $\theta \neq 0$, the integral in
Eq.~\ref{eqn:torque-tot2} is non-zero. To first order in $\theta$, we
can write:
\begin{equation}
J[C(x,\theta)]= J_1 \theta + O(\theta^2),
\label{eqn:J-expansion}
\end{equation}
where $J_1>0$ corresponds to intruders which are unstable to small
perturbations in their angular orientation. Writing an equation for
torque about the center of mass for small angles, and setting it equal
to the moment of inertia, $I_{mom}$, times the angular acceleration,
$\ddot{\theta}$, yields:
\begin{equation}
\ddot{\theta}=\frac{\tau}{I_{mom}}\approx\frac{B_0 J_1}{I_{mom}} v^2 \theta = \Gamma v^2 \theta,
\label{eqn:theta-dynamics}
\end{equation}
where, $\Gamma \equiv \frac{B_0 J_1}{I_{mom}}$. 

\begin{figure}[t!]
\includegraphics[clip, trim=2mm 0mm 2mm 0mm, width=0.9\columnwidth]{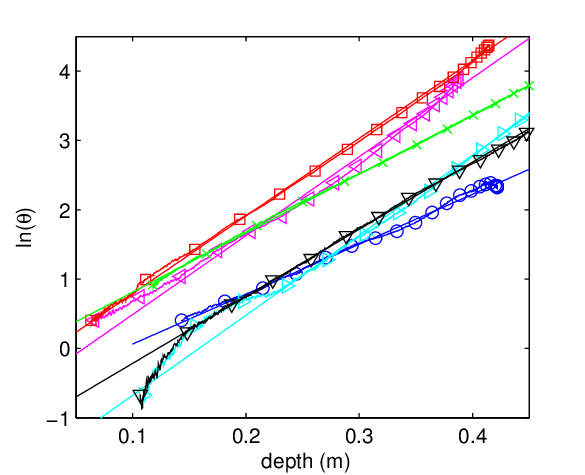}
\caption{(color online.) Plot of the natural log of the angular
  deviation versus depth from single trajectories, where all angular
  deviations are considered positive. The straight lines suggest
  exponential growth for $\theta$ vs. depth; the slope on the semi-log plot
  corresponds to the exponential growth rate, $\lambda_+$, as
  discussed in the text. Different colors and symbols denote different intruders,
  with the same designations as in Figs. \ref{fig:stop-times-and-depths} and \ref{fig:hzcones}.}
\label{fig:rotations}
\end{figure}

Testing this relation nominally requires $\ddot{\theta}$ and $v^2$. As
for the motion of the intruder center of mass, the numerical
computation of each time derivative of $\theta$ obtained from
experiment amplifies the measurement noise. However, examination of
the angular trajectories, vs. $z$ indicates approximately exponential
growth: $\theta(z)\approx \theta_0 e^{\lambda z}$, as shown in
Fig.~\ref{fig:rotations}. Thus, instead of directly testing
Eq.~\ref{eqn:theta-dynamics}, the data suggest that analyzing the
angular orientation as a function of depth, $\theta(z)$ might be a
useful approach. In order to faciliate such analysis, we use product
and chain rules, along with the decelerating force from
Eq.~\ref{eqn:forcelaw}, to write:
\begin{eqnarray}
\ddot{\theta}&=&v^2\frac{d^2\theta}{dz^2}+\frac{dv}{dt} \frac{d\theta}{dz} \nonumber \\
&=&v^2\frac{d^2\theta}{dz^2}+\left(g-\frac{f(z)}{m}-\frac{h(z)}{m}v^2\right)\frac{d\theta}{dz}
\label{eqn:time-to-z}
\end{eqnarray}
Combining this result with Eq.~\eqref{eqn:theta-dynamics} yields:
\begin{equation}
 \frac{d^2\theta}{dz^2}-\left(\frac{h(z)}{m}\right)\frac{d\theta}{dz}-\Gamma \theta= \frac{1}{v^2}\left(\frac{f(z)}{m}-g\right)\frac{d\theta}{dz} .
\label{eqn:theta-dynamics2}
\end{equation}

In the large-velocity regime (i.e., where the velocity-squared force
dominates), the right-hand side of this equation is small, and, if
$h(z)$ is constant, this equation can be easily solved. Examining
Figs.~\ref{fig:trajectories}~and~\ref{fig:hzcones}, we see that
$f(z)\approx mg=2.15$~N for much of the trajectory (recall, the
intruder mass is 0.219 kg), and right side is further reduced for
large velocities by $1/v^2$. Hence, for the moment, we assume that the
right side of Eq.~\ref{eqn:theta-dynamics2} is negligible (we return
later to the validity of this approximation). Replacing $h(z)$ with
$h_0$, we obtain:
\begin{equation}
\theta(z)=\theta_{0+} e^{\lambda_+ z} + \theta_{0-}e^{\lambda_- z},
\label{eqn:zdynamics}
\end{equation}
where $\theta_{0\pm}$ are constants of integration, and
\begin{equation}
\lambda_{\pm}=\frac{h_0}{2m}\pm\sqrt{\left(\frac{h_0}{2m}\right)^2+\Gamma}
\label{eqn:lambda}
\end{equation} 
For $\Gamma>-\left(\frac{h_0}{2m}\right)^2$, the $\lambda_\pm$ are
purely real, with $\lambda_+>0$ and $\lambda_-<0$. In this case, the
collisional model predicts exponential growth in depth of the angle of
rotation, with a growth rate
$\lambda_+=\frac{h_0}{2m}+\sqrt{\left(\frac{h_0}{2m}\right)^2+\Gamma}$. The
quantities $J[C(x,\theta)]$ and $I_{mom}$, which yield $\Gamma$, are
straightforward to calculate from the intruder geometry, and $h_0$ is
already calculated for the intruders used here
(Figs.~\ref{fig:hzcones} and \ref{fig:hvsI}).

\begin{figure}[t!]
\includegraphics[clip, trim=0mm 0mm 0mm 0mm, width=0.87\columnwidth]{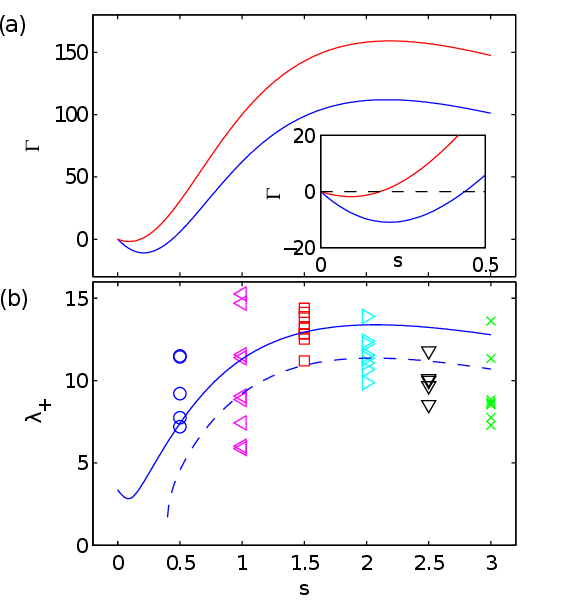}
\caption{(color online.) (a) Plot of $\Gamma$, calculated from
  Eq.~\eqref{eqn:GammaSqr} and preceding equations, with (red, upper curve) and
  without (blue, lower curve) the tip contribution included. As discussed in the
  text, $\Gamma>0$ corresponds to a rotational instability, which
  occurs at $s\approx 0.2$ with the tip included and $s\approx 0.4$
  without the tip included. Thus, $s=0$ intruders should be stable,
  and all other intruders should be unstable, which is consistent with
  data presented here. (b) A plot of all measured values of
  $\lambda_+$ versus the aspect ratio, $s$. Each data point represents
  a trajectory with sufficient angular deviation (i.e., it has at
  least 3000 data points where $\theta>10^\circ$), where wee measure
  the growth rate as shown by the linear fits imposed on each
  trajectory in Figure~\ref{fig:rotations}. Also plotted is the
  prediction for $\lambda_+$ from Eqs.~\eqref{eqn:lambda} and
  \eqref{eqn:GammaSqr} with (solid line) and without (dashed line) the
  contribution from the intruder tip.}
\label{fig:GammaInstability}
\end{figure}

As noted, plots of the natural log of the angle versus depth follow
approximately straight lines, where the slope corresponds to
$\lambda_+$ (Fig.~\ref{fig:rotations}). Leftward and rightward
rotation are both plotted as positive. We measure the slope of these
lines for all trajectories which break symmetry sufficiently (i.e.,
have at least 3000 data points where $\theta>10^\circ$) and which have
initial velocities $v_0>3.5$~m/s. (This ensures that the right-hand
side of Eq.~\ref{eqn:theta-dynamics2} is small for the bulk of the
trajectory). The resulting data for $\theta(z)$ show reasonable
reporducibility for each shape and initial velocity, and clear
variations from one shape to another (bottom panel of
Fig.~\ref{fig:GammaInstability}). Note that no $s=0$ trajectories are
plotted, as they do not show sufficient rotation. Moreover, rotational
stability of the $s = 0$ intruder is expected, since $J_1<0$ (top
panel of Figure~\ref{fig:GammaInstability}). For instance, when $J_1 <
0$, a small perturbation of $\theta$ from zero will lead to behavior
in time that will be roughly oscillatory. For instance, if we freeze
$v^2$, the resulting equation of motion $\ddot{\theta}$ is
oscillatory. However, when $J_1 >0$, the dynamics are predicted to be
saddle-like.

\begin{figure}[th!]
\includegraphics[width=0.8\columnwidth]{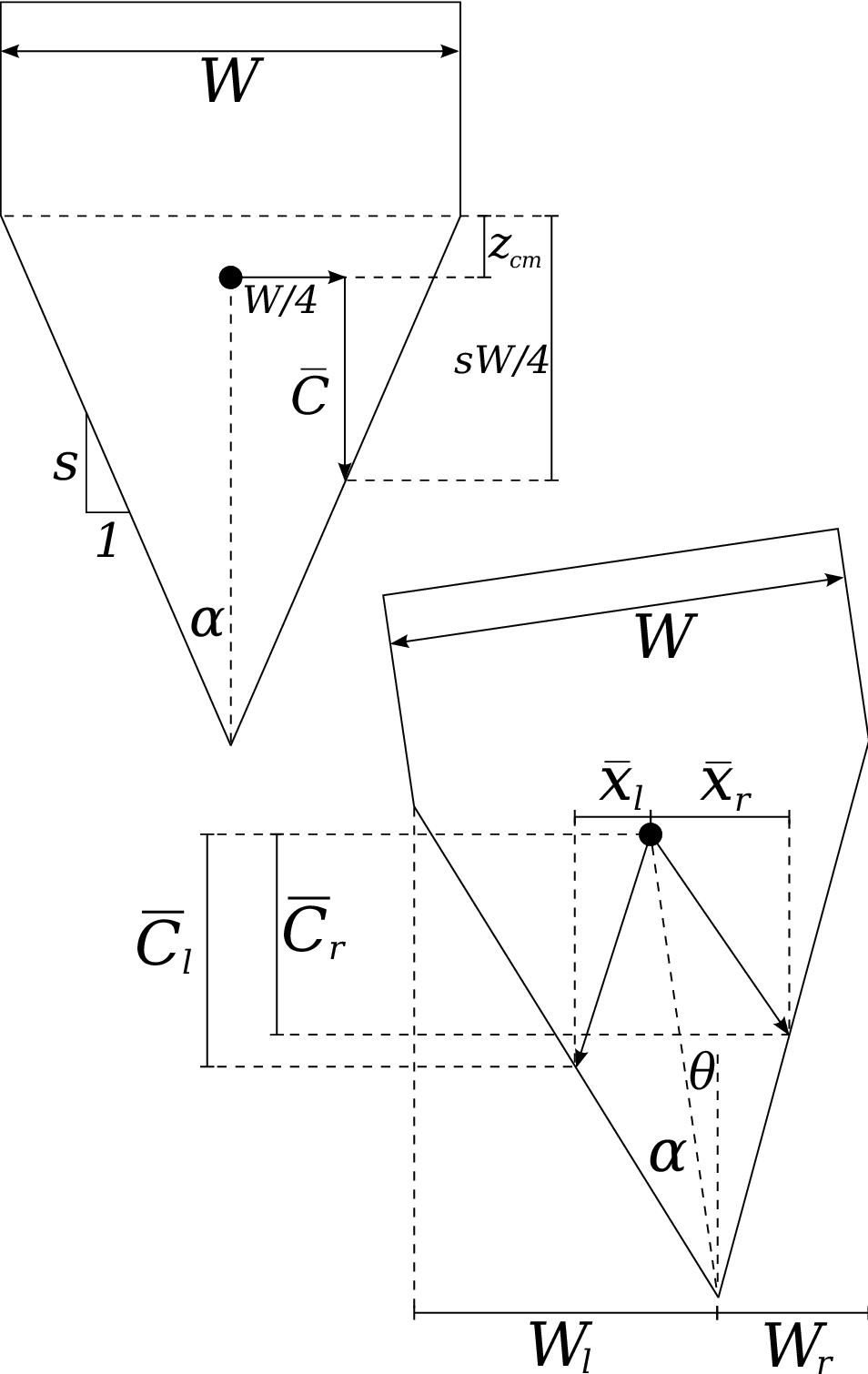}
\caption{A sketch of a triangular-nosed intruder which depicts the
  various quantities used in calculating the torque. See the text for
  details.}
\label{fig:trianglesketch}
\end{figure}

To compare these predictions to the experimental data, we calculate
$\lambda_+$ for the triangular intruders according to equations
\eqref{eqn:torque-tot}-\eqref{eqn:lambda}. Figure~\ref{fig:trianglesketch}
is intended as a visual aid for the following discussion, as it
depicts the geometrical quantities used in this calculation. We break
the integral for $J[C(x,\theta)]$ into two sections for the left and
right sides. We replace the terms involving $C'$ with terms involving
$\alpha$ (the intruder nose angle) and $\theta$; specifically,
$(1+C'^2)^{-1/2}=\sin(\alpha\pm\theta)$ and
$C'(1+C'^2)^{-1/2}=\cos(\alpha\pm\theta)$. This simplifies the
calculation, yielding the torques for the left and right sides,
\begin{eqnarray}
|\vec{\tau}_l| &=& B_0 v^2 W_l \times\nonumber \\ 
		&&\!\!\!\!\left[\bar{C}_l \sin(\alpha + \theta)\cos(\alpha + \theta)-\bar{x}_l \sin^2(\alpha + \theta)\right]\\
|\vec{\tau}_r| &=& B_0 v^2 W_r \times\nonumber \\ 
		&&\!\!\!\!\left[\bar{C}_r \sin(\alpha - \theta)\cos(\alpha - \theta)-\bar{x}_r \sin^2(\alpha - \theta)\right]
\label{eqn:l-and-r-torques}
\end{eqnarray}
where the total torque in
Eqs.~\eqref{eqn:torque-tot}~and~\eqref{eqn:torque-tot2} is then given
by the difference between these two expressions. Here, $z_{cm}$ is the
distance from the nose-tail boundary to the center of mass as shown in
Fig.~\ref{fig:trianglesketch}. Positive $z_{cm}$ is into the
triangular nose, negative $z_{cm}$ is into the rectangular tail,
$W_{l,r}=\frac{W}{2}\sqrt{s^2+1}\sin(\alpha\pm\theta)$ are the
horizontal projections of the left and right sides of the triangular
nose, and
$\bar{C}_{l,r}=(\frac{sW}{4}-z_{cm})\cos\theta\pm\frac{W}{4}\sin\theta$
and
$\bar{x}_{l,r}=\frac{W}{4}\cos\theta\mp(\frac{sW}{4}-z_{cm})\sin\theta$
are the strictly positive vertical and horizontal distances,
respectively, from the center of mass to the midpoint of the sides of
the triangular nose. The expressions for $z_{cm}$ and $I_{mom}$ are
functions of area, $A$, mass, $m$, and width, $W$, although we hold
these quantities constant, as well as functions of the nose aspect
ratio, $s$ which we vary in these experiments:
\begin{eqnarray}
z_{cm}&=&\frac{s^2 W^3}{96 A} + \frac{s W}{4} - \frac{A}{2 W}\\
I_{mom}&=&\left(\frac{A^3}{12 W^2}+\frac{A W^2}{12}\right)-s \left( \frac{W^4}{96} \right)\nonumber \\&& {}+s^2\left(\frac{A W^2}{96} \right)-s^4 \left(\frac{W^6}{9216 A}\right)
\label{eqn:z-and-I-eqns}
\end{eqnarray}
The contribution from the tip must also be included, since the force
there can be substantial, e.g. Fig.~\ref{fig:hvsI}. We model the tip
force, $F_t$, according to the dashed fit line in
Figure~\ref{fig:hvsI}, given by:
\begin{equation}
F_t(s)\approx 0.2 \left(1-e^{-2s}\right)v^2
\label{eqn:tipforce}
\end{equation}
We assume this force always acts in the $-\hat{z}$-direction, which
allows a calculation of the torque from the tip, $\vec{\tau}_t$:
\begin{equation}
|\vec{\tau}_t| = F_t\left(\tfrac{sW}{2}-z_{cm}\right)\sin\theta
\label{eqn:tip-torque}
\end{equation}
Finally, we obtain:
\begin{equation}
\Gamma = \lim_{\theta \rightarrow 0} \frac{1}{\theta} \left(\frac{|\vec{\tau}_l|+|\vec{\tau}_t|-|\vec{\tau}_r|}{v^2 I_{tot}}\right),
\label{eqn:GammaSqr}
\end{equation}
where we consider the small $\theta$ limit.

The top panel of Figure~\ref{fig:GammaInstability} shows a plot of
$\Gamma$ versus aspect ratio, with (red line) and without (blue line)
the contribution from the tip included. Recall from
Eqs.~\eqref{eqn:J-expansion} and \eqref{eqn:theta-dynamics} that
$\Gamma>0$ corresponds to a rotational instability. Thus, with or
without the tip contribution included, the instability occurs between
$s=0$ and $s=0.5$, which is consistent with the experimental data, as
no $s=0$ intruders show substantial rotations. The value for $\Gamma$
is then used to calculate $\lambda_+$ according to
Eq.~\eqref{eqn:lambda}. Note also that there are \textit{no free
  parameters} here, since the values of $B_0\approx 7.6$~Ns$^2$/m and
$F_t$, given by Eq.~\eqref{eqn:tipforce}, are inputs from the
collisional drag measurements shown in Fig.~\ref{fig:hzcones}. The
model prediction for $\lambda_+$ is in good agreement with the
experimental results, as shown in the bottom panel of
Fig.~\ref{fig:GammaInstability}. The dashed line shows the prediction
without the tip contribution, while the solid line includes the tip.
This constitutes the second main result of this paper.

\begin{figure}[t!]
\includegraphics[clip,trim=0mm 4mm 0mm 0mm,width=0.9\columnwidth]{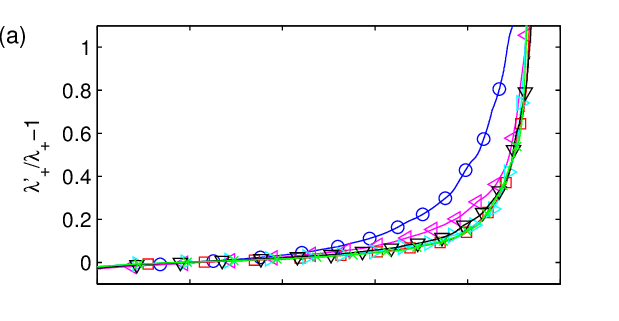} 	      
\includegraphics[clip,trim=0mm 0mm 0mm 4mm,width=0.9\columnwidth]{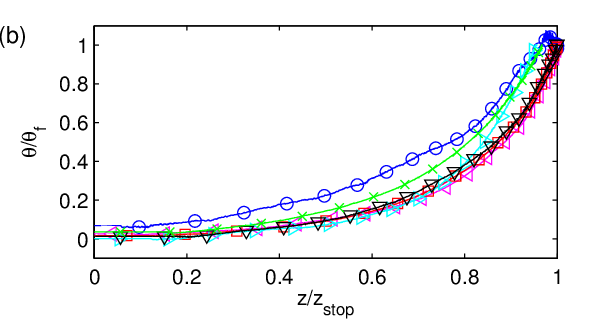}
 \caption{(color online.) (a) Error in the growth rate, $\lambda_+$,
   plotted versus the normalized depth for the intruder trajectories
   shown in Fig.~\ref{fig:rotations}. We calculate this error by
   including the neglected term on the right-hand side of
   Eq.~\eqref{eqn:theta-dynamics}, using the instantaneous velocity
   and depth, which yields $\lambda'_+$. We divide this by $\lambda_+$
   and subtract 1, yielding the error made by the approximation (0
   corresponds to no error, and 1 corresponds to an error of the same
   size as $\lambda_+$). As the intruder slows down, the approximation
   is no longer valid, but this does not occur until very late in the
   trajectory, at approximately 80-90\% of the final depth,
   $z_{stop}$. (b) Plot of the angle versus depth, where the
   angle and depth are both normalized by their final values. In terms
   of total rotation angle, a substantial amount of rotation (roughly
   40\% of the total rotation) happens at the end of the trajectory,
   after the error in $\lambda_+$ begins to grow
   substantially. However, plotted on a semi-log scale, as in
   Fig.~\ref{fig:rotations}, the bulk of the dynamics are clearly
   exponential in depth, with a well defined growth rate.}
 \label{fig:error}
\end{figure}

Concerning the scatter in the data for measured values of $\lambda_+$,
we suggest several possible explanations. First, we neglected the
right-hand side of Eq.~\ref{eqn:theta-dynamics2}. In practice, this
term is associated with a velocity- and depth-dependent correction to
$\lambda_+$ through the coefficient in front of
$\frac{d\theta}{dz}$. The strength of this correction depends on the
velocity and depth at which the bulk of the rotation occurs. However,
Fig.~\ref{fig:error} shows that the error introduced into $\lambda_+$
is small for approximately 80-90\% of the trajectory. Thus, the data
for exponential fits in Fig.~\ref{fig:rotations} are primarily in the
regime where the approximation made here is valid. Second, the
collisional process which is responsible for rotations is stochastic
(in time, in space, and in magnitude). Thus, we expect some
fluctuations, especially when considering a single
trajectory. Previous studies on the dynamics
\cite{Clark2012,Clark2013} showed that the deceleration is highly
fluctuating for individual trajectories, but for long times and many
trajectories, it approaches the average behavior. Third, we assumed
that the force from the tip always pointed directly upwards, but a
horizontal component of the tip force could have a strong effect,
particularly given the relatively large distance from the center of
mass to the tip. Fourth, we assumed that the collisions were
\textit{equally likely} everywhere over the intruder surface. If this
assumption is not strictly true, it would not substantially affect the
velocity-squared drag, but it could have a stronger affect on the
torque. However, on average, the collisional model captures the
overall behavior of exponential growth, as well as the magnitude and
scaling of the growth rate with intruder shape.

\section{Conclusions}
In this paper, we have presented data for the dynamics of
triangular-nosed intruders impinging on a granular bed from above. We
found that the average dynamics are captured well by a macroscopic
force law, Eq.~\eqref{eqn:forcelaw}, and that the magnitude of the
velocity-squared drag force, $h(z)$, depends strongly on intruder
shape, while the static term, $f(z)$, shows very little dependence on
intruder shape. We also observed that intruder rotations become
increasingly significant as the intruder nose is elongated, and that
in such cases, the angle of the intruder grows approximately
exponentially in depth, $\theta(z)\sim e^{\lambda_+ z}$.

Additionally, we have proposed a collisional model for the
velocity-squared drag on an intruder moving through a granular medium,
and we have used experimental data to test this model. The key
component in the model is momentum transfer from the intruder to the
grains. By focusing on intruders with triangular noses, we are able to
systematically explore the effect of intruder shape on the collision
process. We observe experimentally that momentum transfer per unit
surface length is larger at the tip than elsewhere along the sides of
the intruders. By modeling the collision process in terms of both tip
and side contributions, we observe excellent agreement with both the
intruder deceleration and angular orientation. The agreement between
these two \textit{linearly independent} measurements of the mesoscopic
collisional theory serves as an additional confirmation of the basic
assumptions, which are as follows. The velocity squared drag and
angular rotations:
\begin{enumerate}[label=(\alph*),noitemsep,nolistsep]
\item are dominated by intermittent, generally inelastic, collisions;
\item collisions involve grain clusters which can be modeled by as
  having a fixed (mean) mass;
\item collisions occur with equal probability throughout the granular
  material;
\item momentum transfer acts normally inward at the site of collision on the
  intruder surface;
\item there is a disproportionately larger contribution from collisions
  occuring at the tip of the intruder when $s$ is large.
\end{enumerate}

We note several important aspects to the physics that underlie the
model. First, the momentum transfer acts normally at the point of collision,
which essentially says that friction between the grains and intruder is not 
important in this process \cite{Seguin2009}. The shape factors, $I[C(x)]$ and $J[C(x,\theta)]$, 
were derived under this assumption, which is verified by the agreement 
with experimental data (shown in Figs.~\ref{fig:hvsI} and \ref{fig:GammaInstability}).
We note that this is simply for the collisional term, which controls the velocity-squared drag and
the corresponding off-axis rotations.
Another point is that there is a force network that is dynamically
excited by the intruder. If the network were to fail rapidly under the
advance of an intruder, we would not expect to excite `clusters', at
least not in the same way as we observe in the experiments. Also of
importance is the characteristic length of the force network that is
excited by a single event. In Clark et al.~\cite{Clark2012}, we found
that this characteristic length was in the range of ten to a few tens
of grains (note that this is consistent with our estimate of the
cluster size from Fig.~\ref{fig:clustersize}). It is interesting to
contrast the collisional picture of Poncelet with Bagnold scaling~\cite{Bagnold1954} for
shear flow, where shear stresses are expected to vary as $v^2$ due to
inter-particle collisions, but in a much less dense granular
phase. Finally, we note the relevance of the speed of the intruder (at
impact and later) relative to the granular sound speed. In the present
experiments, this ratio is small, no more than 0.02. As the intruder
speed increases relative to the granular sound speed, the present
model scenario may need modification. In particular, when the intruder
speed is high enough, the force signal propagating into the material
may not be able to `run away' from the intruder. In any event, one
might expect the typical cluster size in a collision event to grow as
the impact speed approaches the granular sound speed. 

An interesting question for both experiments and models concerns
intruder dynamics in three dimensions. The fact that a number of 3D
experiments with comparable impact and granular wave speeds also find
$v^2$ scaling for the dynamic part of the drag suggests that a
collisional model of the type developed here could be applied
successfully to 3D. The fact that stresses from collisions can be
determined within $O(1)$ normalization suggests that it may be
possible to use such an approach for applications, such as maximizing
or minimizing the inertial drag or understanding the stability and
dynamics of rotations of granular intruders as a function of
intruder shape. 

This work has been supported by the U.S. DTRA under Grant No.
HDTRA1-10-0021. We very much appreciate additional input from
Dr. L. Kondic, Dr. C. O'Hern, and Dr. W. Losert.

\bibliographystyle{prsty}

\end{document}